# On Shift Sequences for Interleaved Construction of Sequences Sets with low Correlation

N Rajesh Pillai, Yogesh Kumar

*Abstract*— Construction of Signal sets with low correlation property is of interest to designers of CDMA systems. One of the preferred ways of constructing such sets is the Interleaved construction which uses two sequences *a* and *b* with 2-level autocorrelation and a shift sequence *e*. The shift sequence has to satisfy certain conditions for the resulting signal set to have low correlation properties. This article shows that the conditions reported in literature are too strong and give a version which results in more number of shift sequences. An open problem on existence of shift sequences for attaining an interleaved set with maximum correlation value of *v+2* is also taken up and solved.

## I. INTRODUCTION

CONSTRUCTION of signal sets with good correlation properties is of interest to designers of CDMA systems and Spread Spectrum systems. Among the various ways of constructing signal sets Gold [9], Kasami [5], Interleaved [1], Bent [6,7,8], Interleaved Constructions is interesting because of the larger number of sequence sets and good non-linearity properties it provides.

Interleaved signal sets are constructed by taking two sequences with 2-level autocorrelation and combining them under control of a shift sequence *e*. There are conditions [1] that this shift sequence has to satisfy for resulting sets to have the maximum correlation value $\delta \leq 2v+3$. This article shows that the condition[1] is strong. That is, there are shift sequences which do not satisfy the condition but still attain the bound. In [2] a condition was proposed for shift sequences, which if satisfied would lead to construction of signal sets with $\delta \leq v+2$. The existence of such shift sequences was left as an open problem. It is shown in this paper that shift sequences satisfying the condition given do not exist.

This paper is organized as follows: In Section II, we give some preliminary concepts about sequences that we use in this paper (for more theory on sequences refer [2,10]). In Section III, interleaved construction and the condition on shift sequence are given along with a concrete example. We also give the modified version of the condition and argue about its correctness. In Section IV we state the open problem from [2], and then solve it.

## II. PRELIMINARIES

In this section, we present some preliminaries about sequences. We will use the following notation throughout the paper.
- *p*, a prime integer.
- $Z_v$ represents the ring of integers modulo *v*.
- $F_q$ =GF(q), a field with q elements.
- $F_p^n$ = $\{(x_0, x_1, x_2,...,x_{n-1}) \mid x_i \in F_p \}$, a vector space over $F_p$.
- $a = \{a_i\}$, a sequence over $F_p$. If *a* is a periodic sequence with period *v*, then we denote $\mathbf{a} = (a_0, a_1, a_2, \ldots, a_{v-1})$,

Let $a = \{a_i\}$ be a *p*-ary sequence over $F_p$. The left shift operator *L* on *a* is defined as $L(a) = a_1, a_2, a_3, \ldots$, i.e. the left shift operator *L* when applied to a sequence will shift the sequence to left by one position.

For any i > 0, $L^i(a) = a_i, a_{i+1}, a_{i+2}, \ldots$, is said to be phase shift of *a*. Two periodic sequences *a* and *b* are shift equivalent if there exists an integer k such that $a_i = b_{i+k}$, $\forall$ i ≥ 0. Otherwise they are said to be shift distinct.

*Correlation*

Let $a = (a_0, a_1, a_2,..., a_{v-1})$ and $b = (b_0, b_1, b_2, \ldots, b_{v-1})$ be two sequences over $F_p$ with period *v*, their crosscorrelation function

$$C_{a,b}(\tau) = \sum_{i=0}^{v-1} \omega^{a_i - b_{i+\tau}} \quad \tau = 0,1,2...$$

Where $\omega = e^{2\pi i/p}$, a *p*-th primitive root of unity.

If $b = a$, then $C_{a,a}(\tau)$ is called an autocorrelation function of *a*, denoted by C(τ). If

$$C(\tau) = \begin{cases} v & \text{if } \tau \equiv 0 \mod v \\ -1 & \text{otherwise} \end{cases}$$

then we say that the sequence *a* has a (ideal) 2-level autocorrelation function.

If *a* and *b* are shift distinct and $|C_{a,b}(\tau)| \leq \delta$ for any τ, where δ is a constant close to $\sqrt{v}$, the square root of *v*. then we say that *a* and *b* have low crosscorrelation function.

Let $s_j = (s_{j,0}, s_{j,1}, s_{j,2}, \ldots, s_{j,v-1})$, $0 \leq j < r$ be *r* shift distinct sequences over $F_p$ of period *v*. let S = $(s_0, s_1, s_2, \ldots, s_{r-1})$ and $\delta = \max |C_{si,sj}(\tau)|$ for any $0 \leq \tau < v$, $0 \leq i, j < r$, where $\tau \neq 0$ if $i = j$.

N Rajesh Pillai and Yogesh Kumar are with Scientific Analysis Group, Metcalfe House Complex, Delhi, INDIA, ( email: nrpillai@yahoo.com)

yogesh_lather@)yahoo.co.in )

*Interleaved Sequences*

Let $u = (u_0, u_1, u_2, ..., u_{st-1})$ be a sequence over $F_p$ of period $st$, where both $s$ and $t$ are not equal to 1. We can arrange the elements of the sequence **u** into an s by t array as follows:

$$A = \begin{pmatrix} u_o & u_1 & \cdots & u_{t-1} \\ u_t & u_{t+1} & \cdots & u_{2t-1} \\ \vdots & \vdots & & \vdots \\ u_{(s-1)t} & u_{(s-1)t+1} & \cdots & u_{(s-1)t+t-1} \end{pmatrix}$$

If each column vector of the above array is either phase shift of a sequence over $F_p$, say *a*, of period *s*, or zero sequence, then we say that *u* is an (s, t) interleaved sequence over $F_p$ [3]. We also say that A is the matrix form of *u*. let $A_j$ be $j^{th}$ column vector of A, then $A = (A_0, A_1, A_2,..., A_{t-1})$. We call the $A_j$ column sequences of *u* (or component sequences of *u*). According to the definition, $A_j$ is a transpose of either $L^{e_j}(a)$, or ( 0, 0, 0, ... , 0). But we omit the transpose and write $A_j = L^{e_j}(a)$, $0 \leq j < t - 1$ where we denote $e_j = \infty$ if $A_j = (0, 0, 0, ... , 0)$. We say **u** is an (s, t) interleaved sequence associated with (*a, e*). and $e = (e_0, e_1, e_2, ... , e_{t-1})$ is called a shift sequence of *u*.

In interleaved construction of signal sets with low correlation, we start with sequences *a*, *b* of period *v* with 2-level autocorrelation, and a shift sequence *e*. First an interleaved sequence *u* is constructed using *a* and *e*. The signal set $S = \{u+L^i(b) \mid 0 \leq i < v \}$ has low correlation properties provided sequence *e* satisfies certain conditions.

In order to conveniently study the conditions on the shift sequence *e*, we extend it from $(e_0, e_1, e_2, ... , e_{v-1}) \in Z_v^v$ to $(e_0, e_1, e_2, ..., e_{2v-1}) \in Z_v^{2v}$ by defining $e_{v+j} = 1 + e_j$, $0 \leq j < v$

*Lemma* [1]: Let $s_h$ and $s_k$ be two sequences in *S*. Let $\tau = rv + s$ with $0 \leq s < v, r \geq 0$. then $C_{h,k}(\tau)$, the cross-correlation between $s_h$ and $s_k$, can be computed by

$$C_{h,k}(\tau) = \sum_{j=0}^{v-1} C_a(t_j)\omega^{d_j}$$

Where $d_j = b_{h+j} - b_{k+s+j}$ and $t_j = e_{j+s} - e_j + r$.
Note: we use $e_{v+j} = e_j + 1$ where $j \geq 0$.

Following the approach used in proof of Theorem 2 [1] it can be shown that cross-correlation will be

$C_{h,k}(\tau) = - C_a(r) \in \{1, -v\}$ when $s = 0$,

and when $s \neq 0$, the maximum magnitude of cross-correlation is given by

$1 + (v+1)($ Number of times $t_j \equiv 0 \bmod v)$

For a given pair of sequences, for a given shift, knowing the number of times $t_j \equiv 0 \bmod v$, will help in determining the magnitude of cross-correlation

Then we say that the set S has low cross-correlation. We call *S* a (*v, r, δ*) signal set [4].

## III. SHIFT SEQUENCES FOR INTERLEAVED CONSTRUCTION

*Interleaved Construction*

This construction gives signal sets of the form $(v^2, v + 1, 2v + 3)$ where *v* is either prime or of form $2^n-1$.

Given two binary sequences *a, b* of period *v* with 2-level autocorrelation, let $e = (e_0, e_1, e_2, ..., e_{v-1})$, $e_i \in Z_v$ be the shift sequence. *u* is interleaved sequence whose *jth* column sequence is $L^{e_j}(a)$

Signal set $S = \{S_j \mid j = 0, 1, 2, ..., v - 1\} \cup \{u\}$

Where $S_j = u + L^j(b)$, $0 \leq j < v$.

If the shift sequence *e* satisfies [1]

$\mid \{e_j - e_{j+s} : 0 \leq j < v - s\} \mid = v - s$, for all $s$, $1 \leq s < v$ ——— (A)

then the signal set *S* is of type $(v^2, v + 1, 2v + 3)$.

To show that the condition (A) is too strong, we give a shift sequence *e* which does not satisfy (A) but gives a $(v^2, v + 1, 2v + 3)$ signal set when combined with sequences with 2-level autocorrelation.

*Example*

For *v*=7, *e*=(0 0 1 0 6 3 5) with *a*=(1 0 0 1 1 1 0 ) *b*=(1 0 0 1 0 1 1) gives a (49,8,17) signal set. It can be verified that *e* does not satisfy (4). In particular,

For $s=1$, $\mid \{e_j - e_{j+s} : 0 \leq j < v - s\} \mid = \mid \{0,6,1,1,3,5\} \mid = 5$
Where as $v - s = 7-1 = 6$

A more general condition on *e* will be in terms of the extended shift sequence where $e_k = e_{k-v} + 1$ for $k \geq v$

The shift sequence *e* should be such that for all *s*, $1 \leq s < v$ any element in the multiset, $\{e_j - e_{j+s} : 0 \leq j < v \}$ has at most two repetitions ——— (B)

(A multiset is a collection of elements where repetition is allowed.)

All the proofs for sequence set properties in [1] hold without any modification for this condition. In fact in the proof of the Theorem 2 in [1] it is shown that for the correlation properties, the requirement on *e* sequence is that the number of times

$t_j \equiv 0 \bmod v$

is at most 2. Or in other words after substituting for $t_j$, for fixed *r* and *s*, the number of times the equation

$e_j - e_{j+s} \equiv r \bmod v$, $0 \leq j < v$

is satisfied is at most 2 which is same as condition (B)
It can be verified that shift sequence given in example above satisfies condition (B) and hence gives a (49, 8, 17) set.
It is possible to show that condition (A) implies condition (B)

*Lemma*: Condition (A) implies condition (B)
*Proof*: Suppose a shift sequence *e* satisfies condition (A) but not (B). Then for some *s*, there exists at least three values of *j* (say $j_1, j_2$ and $j_3$) for which $e_j - e_{j+s}$ evaluates to the same value modulo *v*.

For convenience let $L_1$ denote the sequence of values ( $e_j - e_{j+s}$ : $0 \leq j < v - s$) and $L_2$ denote ( $e_j - e_{j+s}$ : $v-s \leq j < v$ ). Then we have two cases-

Case 1: At least two of the $j_i$'s are in $L_1$ which mean a repetition of values in $L_1$ for shift $s$, so condition (A) is not satisfied which contradicts our assumption

Case 2: At most one of the $j_i$'s is in $L_1$. This means at least two of them are in $L_2$. That is
$|\{ e_j - e_{j+s} : v-s \leq j < v \}| < s$
$\Rightarrow |\{ e_j - e_{j+s-v} +1 : v-s \leq j < v \}| < s$   since $e_{j+s} = e_{j+s-v} + 1$
$\Rightarrow |\{ e_j - e_{j+s-v} : v-s \leq j < v \}| < s$
Substituting $k$ for $j+s-v$ we have
$|\{ e_{k+v-s} - e_k : 0 \leq k < s \}| < s$
Substituting $v-s$ by $s'$ we have
$|\{ e_{k+s'} - e_k : 0 \leq k < v-s' \}| < v-s'$
$\Rightarrow |\{ e_k - e_{k+s'} : 0 \leq k < v-s' \}| < v-s'$
which again contradicts our starting assumption that (A) was satisfied.
This proves that (A) $\Rightarrow$ (B).

## IV. Non-existence of Shift Sequences Satisfying Stronger Condition

A research problem was suggested in [2] (Problem 9 of Chapter 10). The problem statement is given below

*Problem:* For an interleaved signal set with parameters $(v^2, v+1, 2v+3)$, the crosscorrelation of any pair of the sequences in the signal set or the out-of-phase autocorrelation of any sequence in the signal set will be reduced to the set $\{1, -v, v+2\}$ if the shift sequence satisfies the following condition:
for all $1 \leq s < v$,
$|\{e_j - e_{j+s} : 0 \leq j < v - s\}$
$\qquad \cup \{e_{v-s+j} - e_j - 1 : 0 \leq j < s\}| = v$  ———— (1)
Does such a vector $e$ exist?

The condition (1) can be equivalently written as
For all $1 \leq s < v$,
$|\{e_k - e_{k+s} : 0 \leq k < v - s\}$
$\qquad \cup \{e_k - e_{k+s-v} - 1 : v-s \leq k < v\}| = v$  ———— (2)

We will show that such sequences do not exist for $v > 2$. Suppose we have such an $e$. Pick any $s$, $1 \leq s < v$, and label the differences as $a_k$'s  We have
$(e_k - e_{k+s}) \equiv a_k \mod v$, $k = 0, 1, 2, 3, ..., v-s-1$
$(e_k - e_{k+s-v} - 1) \equiv a_k \mod v$, $k = v-s, ..., v-1$
Where $a_i$s are the shifted difference modulo $v$ for the shift $s$. Adding all the equations we get

$$\sum_{k=0}^{v-1} e_k - \sum_{k=0}^{v-s-1} e_{k+s} - \sum_{k=v-s}^{v-1} e_{k+s-v} - s \equiv \sum_{k=0}^{v-1} a_k \mod v$$

The condition all $a_k$s are unique modulo $v$, $0 \leq a_k < v$ with $k = 0, 1, 2, ..., v-1$ means $\{a_k : 0 \leq k < v\} = \{0, 1, ..., v-1\}$. So the RHS $\sum_{k=0}^{v-1} a_k = v(v-1)/2 \equiv 0 \mod v$  for odd $v$
$\equiv v/2$  for even $v$
irrespective of the value of $s$. Whereas, on the LHS the $e_k$s cancel out and the equation becomes $-s \equiv v(v-1)/2 \mod v$, for all $s$, $1 \leq s < v$.

Which is not possible as RHS is a fixed value for all the shifts $s$ (even though individual $a_j$s will be different for each $s$, the set of $a_i$s will always be $\{0, 1, 2, ..., v-1\}$) and the value on LHS is changing.
Hence the assumption that all $a_i$s are unique is false, ie. No such $e$ vector exists for $v > 2$.
When $v=2$, there is only one shift $s=1$ to be checked for distinctness of the differences so the contradiction above is avoided. The shift sequence $(0,1)$ satisfies the condition. However in this case $v+2 = v^2$ and the benefits of low correlation are not realised.

## V. Summary

Gong [1] has given condition for shift sequences so that the resulting interleaved sequences have correlation bounded by $2v+3$. This condition was shown to be too strong with the help of a concrete example that it was sufficient but not necessary for obtaining correlation bound of $2v+3$. The actual condition implicit in the proof of Theorem 2 of [1] in terms of the extended shift sequence was stated. A proof that Gong's condition implies the condition on extended sequence was also given.

A research problem (Problem 9, Chapter 10) [2] about existence of shift sequences for obtaining a correlation bounded by $v+2$ was considered and solved.

## VI. Acknowledgement

The authors wish to thank Dr. P. K. Saxena, Dr. S. S. Bedi for encouraging us to work on this problem. They also wish to thank Roopika Chaudhary, Indivar Gupta and Jasbir Singh for various discussions.

## References

[1] G. Gong, "New designs for signal sets with low crosscorrelation, balance property and large linear span: GF(p) case", *IEEE Trans. on Inform. Theory,* Vol.48, No.11, November 2002, pp. 2847-2867.
[2] S.W. Golomb and G. Gong, "*Signal Design with Good Correlation: for Wireless Communications, Cryptography and Radar Applications*", Cambridge University Press, 2005.
[3] G. Gong, "Theory and applications of q-ary interleaved sequences", *IEEE Trans. on Inform. Theory,* Vol .41, No. 2, March 1995, pp. 400-411.
[4] John G. Proakis, "*Digital Communications*", McGraw-Hill, Inc., the third edition, 1995.
[5] J.S. No and P.V. Kumar, "A new family of binary pseudo-random sequences having optimal periodic crosscorrelation properties and large linear span", *IEEE Trans. of Inform.Theory*, Vol.35, No.2, March 1989, pp. 371-379.
[6] J. D. Olsen, R.A.Scholtz and L.R. Welch, "Bent-function sequences", *IEEE Trans. on Inform. Theory,* vol. 28, No. 6, Nov 1982, pp. 858-864.
[7] P. V. Kumar and R. A. Scholtz, "Bounds on the linear span of bent sequences", *IEEE Trans. Inform. Theory*, Vol. IT-29, No. 6, Nov. 1983, pp. 854 - 862.
[8] A. Lempel and M. Cohn, "Maximal families of bent sequences", *IEEE Trans. Inform. Theory,* Vol. 28, No. 6, Nov. 1982, pp. 865-868.
[9] R. Gold, "Optimal binary sequences for spread spectrum multiplexing", *IEEE Trans. Inform. Theory*, Vol. 13, No. 5, Oct. 1967, pp. 619-621.
[10] S. Golomb, *Shift Register Sequences*, Oakland, CA: Holden-Day, 1 1967. Revised edition: Laguna Hills, CA; Aegean Park Press, 1982.